\documentclass[traditabstract,a4paper]{aa}
\usepackage{txfonts} 

\usepackage{color}

\usepackage{natbib}
\bibpunct{(}{)}{;}{a}{}{,} 

\usepackage{graphicx}
\usepackage{fixltx2e}

\usepackage{longtable}
\usepackage{multirow}
\usepackage{overpic}
\usepackage{array}

\definecolor{RoyalPurple}{cmyk}{0.75,0.9,0,0.1}
\definecolor{MyBlue}{rgb}{0.0025,0.1125,0.2}
\usepackage[breaklinks, colorlinks, citecolor=blue, linkcolor=MyBlue, urlcolor=RoyalPurple, 
            colorlinks=true, linkcolor=blue, debug, baseurl=' ']{hyperref}

\graphicspath{{Figures/}}

\begin{document}

\title{Maturity of lumped element kinetic inductance detectors for space-borne instruments in the range between 80 and 180~GHz}

\author{
A.~Catalano \inst{1,2}
\and
A.~Benoit \inst{2}
\and
O.~Bourrion \inst{1}
\and
M.~Calvo \inst{2}
\and
G.~Coiffard \inst{3}
\and
A.~D'Addabbo \inst{4,2}
\and
J.~Goupy \inst{2}
\and
H.~Le Sueur \inst{5}
\and
J.~Mac\'{\i}as-P\'erez \inst{1}
\and
A.~Monfardini \inst{2,1}
} 

\offprints{A. Catalano - catalano@lpsc.in2p3.fr} 

\institute{LPSC, Université Grenoble-Alpes, CNRS/IN2P3, 
\and Institut N\'eel, CNRS, Universit\'e Joseph Fourier Grenoble I, 25 rue des Martyrs, Grenoble, 
\and Institut de Radio Astronomie Millim\'etrique (IRAM), Grenoble,
\and LNGS - Laboratori Nazionali del Gran Sasso - Assergi (AQ),
\and Centre de Sciences Nucl\'eaires et de Sciences de la Mati\`ere (CSNSM), CNRS/IN2P3, bat 104 - 108, 91405 Orsay Campus}

\abstract
{This work intends to give the state-of-the-art of our knowledge of the performance of lumped element kinetic inductance detectors (LEKIDs) at millimetre wavelengths (from 80 to 180~GHz). We evaluate their optical sensitivity under typical background conditions that are representative of a space environment and their interaction with ionising particles. 
Two LEKID arrays, originally designed for ground-based applications and composed of a few hundred pixels each, operate at a central frequency of 100 and 150~GHz ($\Delta \nu / \nu$ about 0.3). Their sensitivities were characterised in the laboratory using a dedicated closed-cycle 100~mK dilution cryostat and a sky simulator, allowing for the reproduction of realistic, space-like observation conditions.
The impact of cosmic rays was evaluated by exposing the LEKID arrays to alpha particles ($^{241}$Am) and X sources ($^{109}$Cd), with a read-out sampling frequency similar to those used for Planck HFI (about 200~Hz), and also with a high resolution sampling level (up to 2~MHz) to better characterise and interpret the observed glitches. In parallel, we developed an analytical model to rescale the results to what would be observed by such a LEKID array at the second Lagrangian point. We show that LEKID arrays behave adequately in space-like conditions with a measured noise equivalent power (NEP) close to the CMB photon noise and an impact of cosmic rays smaller with respect to those observed with Planck satellite detectors.}


\titlerunning{Lumped Elements Kinetic Inductance Detectors maturity for Space Missions}

\maketitle

\section{Introduction}\label{intro}
We live in the precision cosmology era as a result of technological improvements in millimetre-wave experiments. These include space-borne experiments, such as COBE~\citep{cobe}, WMAP~\citep{wmap}, Herschel~\citep{hershel}, Planck~\citep{planck1};  ground-based experiments include, for example POLARBEAR~\citep{2014ApJ...794..171T}, BICEP~\citep{2014ApJ...792...62A}); and balloon-borne experiments encompass, for example, BOOMERanG~\citep{2006ApJ...647..799M}, MAXIMA~\citep{2003NewAR..47..727J}, Archeops~\citep{2004A&A...424..571B}, Dasi~\citep{2002Natur.420..772K}, QUaD~\citep{2002Natur.420..772K}, and ACT~\citep{2013AAS...22110502H} to cite the most important. In particular, the last generation space-borne experiment Planck, after five full sky surveys has shown the most accurate picture of the primordial Universe in temperature and in polarisation~\citep{cosmo}. By contrast, it has revealed a wide range of new systematic errors pointed out thanks to the high sensitivity of the bolometers mounted in the High Frequency Instrument (HFI). Despite the fact that Planck produced precise cosmic microwave background (CMB) polarisation maps, the Planck mission was not conceived as the ultimate instrument for CMB polarisation measurements. Most probably, the B-modes of CMB polarisation~\citep{HuWhite}, which are not sourced by standard scalar type perturbations, will be only marginally detected by Planck. 
For this reason, new proposed space missions, such as CORE+~\citep{core}, PIXIE ~\citep{kogut2011}, and LiteBIRD~\citep{hazumi} are under study. To accomplish this new challenge, it is necessary to improve the overall noise equivalent power (NEP) of the instrument by more than one order of magnitude with respect to the Planck performances (from $10^{-17}$~WHz$^{-1/2}$ to $10^{-18}$~WHz$^{-1/2}$)\citep{2014JCAP...02..006A,2011arXiv1102.2181T}. This can be achieved by increasing the focal plane coverage, using thousands of background limited instrument performance (BLIP) contiguous pixels. In parallel, the control and mitigation of systematic effects has to be taken into account as a design constraint for future generation detector arrays for space applications. In particular, the impact of cosmic rays on detectors and, as a consequence, on the final quality of the data, has been shown to be one of the key points for previous far-infrared space missions. 
\\
In this context, lumped element kinetic inductance detectors (LEKIDs) have now reached a maturity that is adequate to be competitive with other technologies for next generation millimetre and sub-millimetre wave experiments. This was first demonstrated with the use of such detectors in ground-based experiments, in particular in the New IRAM KID Array (NIKA) instrument~\citep{adam2014,catalano_nika2014, monfardini2012}. This kind of detector exhibits background-limited performance under ideal, i.e. single pixel read-out, cold blackbody, electrical measurement, conditions \citep{Mauskopf}, and NEP in the high 10$^{-17}$ range under the 5-10~pW per pixel load typical of the NIKA camera at 150~GHz \citep{2014JLTP..176..787M}.
In this work, we begin using two LEKID arrays consisting of hundreds of pixels, observing at central frequency of 100 and 150~GHz originally optimised for the NIKA ground-based instrument, to measure their sensitivity and the impact of ionising particles under a background representative of a space environment.  

\begin{table*}
\begin{center}
\begin{tabular}{ccc}
\hline
\hline
 &  3~mm array  & 2~mm array \\
\hline \hline
Valid pixels [\#] &  132 & 132\\
Pixel size [mm] &  2.3 & 2.3 \\
Film & Titanium-Aluminium bi-layer  &  Aluminium \\
Film thickness [nm] & 10-25 &  18 \\
Silicon wafer thickness [$\mu$m] & 525 &  300 \\
Transition critical temp [K] & 0.9 & 1.5 \\
Frequency cut-off [GHz] & 65 & 110 \\
Polarised sensitive detectors &   non & non \\
Optical background [pW] &  0.3 & 0.5 \\
Angular size [F$\lambda$]  &  0.75 & 0.75 \\
Overall optical efficiency [\%]  &  30 & 30 \\
\hline \hline
\end{tabular}
\end{center}
\caption{Characteristics of the two LEKID detector arrays.}
\label{tab:1}
\end{table*}

This paper is structured as follows: Sec.\ref{previous} describes the impact of cosmic rays in previous space missions. Sec.\ref{LEKID} introduces the LEKIDs that were used for this study. In Sec.\ref{perf_test}, we describe the laboratory tests that permitted the characterisation of the sensitivity performance and the systematic errors induced by ionising particles. Finally, in Sec.\ref{sim_toi} we simulate timestreams of LEKID data in the absence of sky signals to derive the impact of cosmic rays on LEKIDs in space.

\section{Cosmic rays impact in previous far-infrared space missions}\label{previous}

Cosmic rays (CRs;~\cite{Mewaldt2010, Leske2011}) at a typical orbital configuration (i.e. low Earth orbit or second Lagrangian point) are essentially composed of massive particles: about 88~\% protons, 10~\% alpha particles, 1~\% heavier nuclei, and less than 1~\% electrons.
The energy spectrum of CRs peaks around 200~MeV, corresponding to a total proton flux of 3000\,-\,4000\,particles\,m$^{-2}$\,sr$^{-1}$\,s$^{-1}$\,GeV$^{-1}$. This flux is dominated by galactic CRs in particular in periods of low solar activity. The solar wind decelerates the incoming particles and stops some of those with energies below about 1~GeV. Since the strength of the solar wind is not constant because of changes in solar activity, the level of the CRs flux varies with time~\citep{Mewaldt2010}. 

The impact of CRs on the detectors time-ordered data has been observed in previous far-infrared space missions that used bolometers. For example, glitches in the COBE-FIRAS data were identified to be caused by cosmic-particle hits on the detectors as they were not correlated to the pointing of the
mirrors~\citep{Fixsen}. The number of glitches observed for this experiment was sufficiently small and their removal was not a major problem. Glitches have been also identified in the Herschel Space Observatory both in the SPIRE (spiderweb bolometers operated at temperatures close to 0.3~K;~\cite{Griffin}) and PACS (high-impedance bolometers at 0.3~K;~\cite{2012arXiv1207.5597H}) instruments. Two types of glitches have been observed in the SPIRE detector timelines: large events and smaller co-occurring glitches, both associated with the impact of cosmic rays on the arrays. In the PACS instrument, the energy deposited by cosmic rays raised the bolometer temperature by a factor ranging from 1 to 6~\% of the nominal value. Moreover, 25~\% of the hits depositing energy on the bolometer chips affect the adjacent pixels.

For the purposes of this paper, in particular, we refer to our results on  the impact of CRs observed by HFI of Planck. The Planck satellite\footnote{$http://www.esa.int/Planck$} observed the sky between August 2009 and August 2013 in the frequency range from 30~GHz to 1~THz~\citep{planck1}. This comprised a telescope, a service module, and two instruments: the HFI and the Low Frequency Instrument (LFI). The HFI operated
with 52 high-impedance bolometers cooled to 100~mK in a range of frequencies 
between 100~GHz and 1~THz. In the CMB channels  (between 100~GHz and 300~GHz), the HFI sensitivity per pixel reached exceptional performance corresponding to a NEP of between $1-2 \cdot 10^{-17}$~$WHz^{-1/2}$ ~\citep{Planck2011perf}. By contrast, the HFI detectors exhibited a strong coupling with CR radiation, which produces transient glitches in the raw time-ordered information (TOI) with a rate of about 1~Hz and a template that can be fitted by a sum of various first-order, low-pass filters with a main time constant between 4 and 10~ms plus low time constants up to 2~s. Flight data from HFI and ground tests provided strong evidence that the dominant family of glitches observed in flight are due to CR absorption by the silicon substrate on which the HFI detectors reside~\citep{catalano_glitch, Planck2013glitch}. Glitch energy is propagated to the thermistor by ballistic\footnote{Ballistic conduction is the unimpeded flow of energy that carries charges over large distances within a material.} phonons, with non-negligible contribution by thermal diffusion. The average ratio between the energy absorbed per glitch in the silicon wafer and that absorbed in the bolometer is, in this specific case, about a suppression factor of 650 \citep{catalano_glitch}.

\begin{figure}[b!]
\begin{center}
\includegraphics[width=8cm, , keepaspectratio]{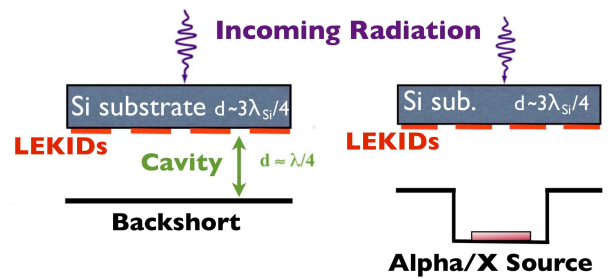}
\end{center}
\caption{Set-ups adopted for sensitivities measurements (left panel) and for glitch characterisation (right panel).}
\label{fig:setup_1}
\end{figure}

\begin{figure*}
\begin{center}
\includegraphics[width=8cm, , keepaspectratio]{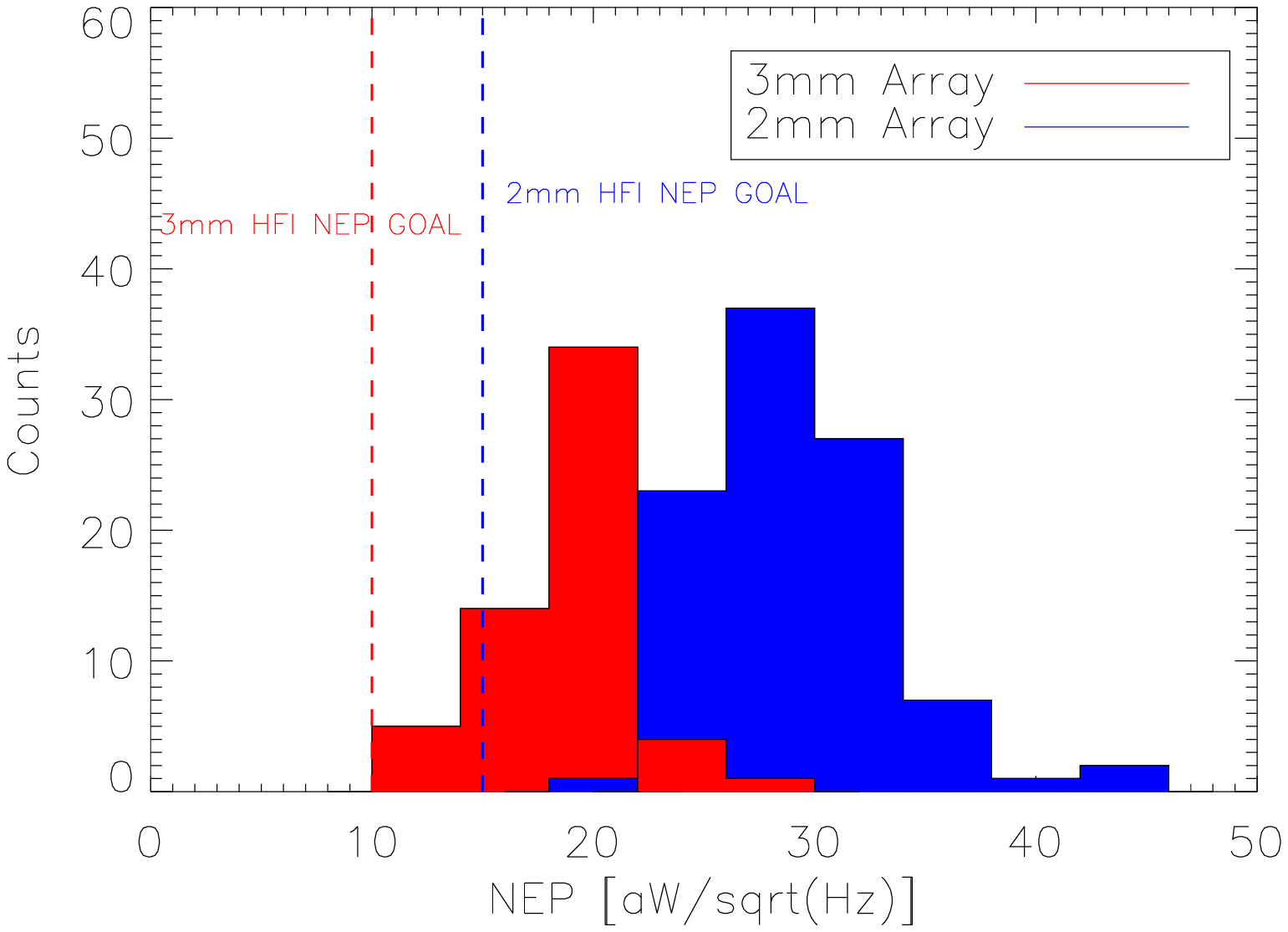}
\includegraphics[width=8cm, , keepaspectratio]{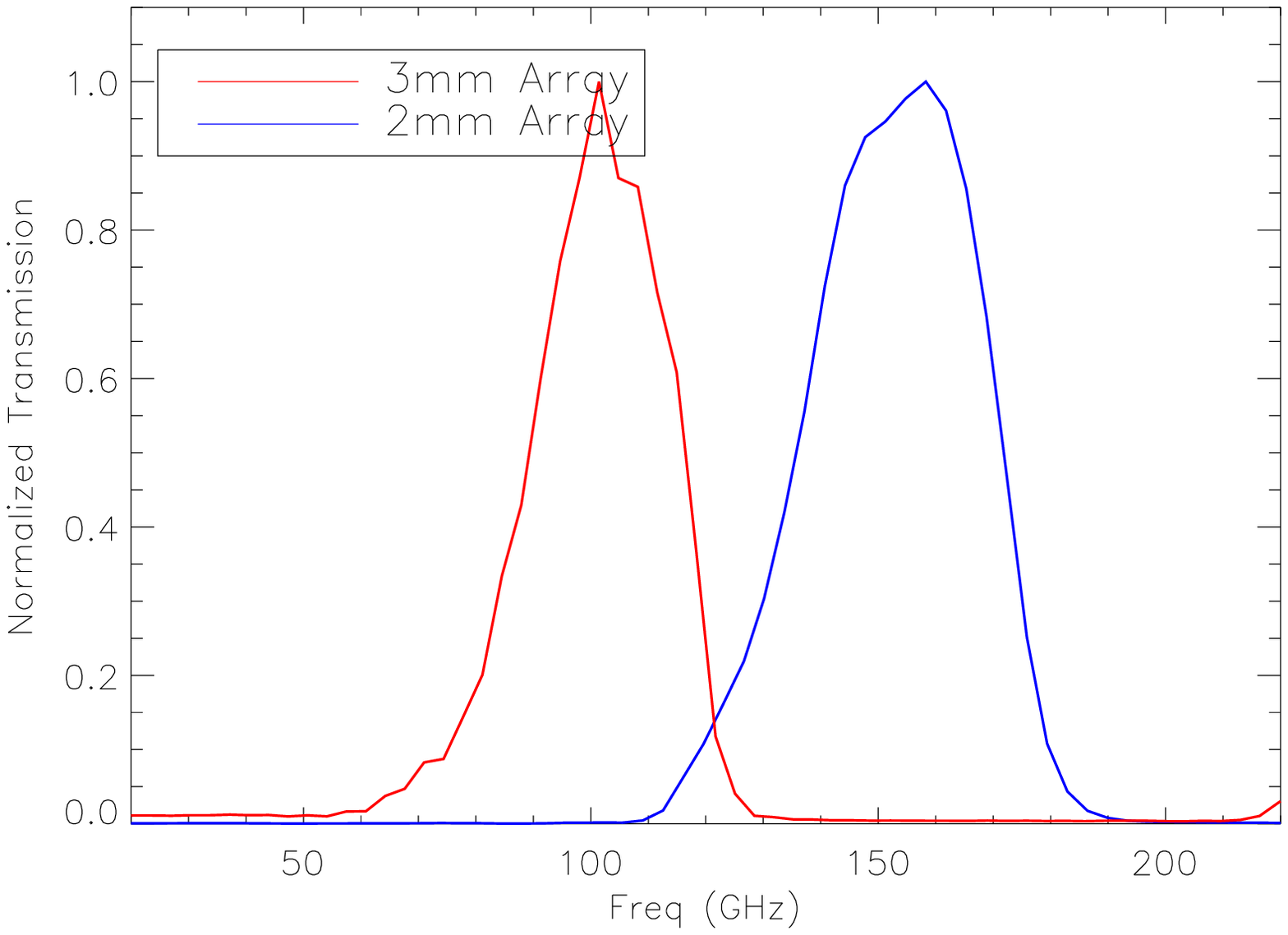}
\end{center}
\caption{Left panel: distribution of pixel sensitivities (red for 3~mm array, blue for 2~mm array) compared to the reference goals of Planck HFI detectors. Right panel: averaged normalised spectral responses (red for 3~mm array, blue for 2~mm array). The spectrum bandwidth ($\Delta \nu / \nu = \frac{FWHM}{\nu_0}$) is 0.28 and 0.32 for  the 3~mm and  2~mm arrays respectively.}
\label{fig:nep_kid}
\end{figure*}

\section{Lumped element kinetic inductance detectors}\label{LEKID}

  For a complete review of the KID theory, we suggest ~\cite{zmudinas} and \cite{doyle}. Here we briefly describe the LEKID principle. Arrays are based on a series of LC resonators fabricated from superconducting strips that are weakly coupled to a 50~$\Omega$ feed-line. The absorbed photons change the Cooper pairs (lossless carries) density producing a change in both the resonant frequency and the quality factor of the resonator. This device acts directly as the absorber of photons at hundreds of GHz. 
We adopted dual polarisation LEKID designed based on a Hilbert fractal pattern for both of the arrays tested in this work~\citep{roesch}. Each pixel is composed of a meander inductor and an interdigitated capacitor. The geometrical characteristics of the pixels are presented in Tab~\ref{tab:1}.
The 2~mm array is made of 132 pixels obtained from 18~nm aluminium film on a 300~$\mu$m HR silicon substrate. In the case of the 3~mm array, since the frequency range below 110~GHz is not accessible using aluminium thin films because of the superconducting gap cutoff, we used bi-layer titanium-aluminium films.
A more detailed explanation of this innovative solution is presented in ~\citep{catalano_3mm}.

\section{Performance testing}\label{perf_test}

The LEKID arrays are cooled at a base temperature of 100 mK in a closed-cycle $^3$He - $^4$He dilution cryostat designed for optical measurements. The cryostat hosts two independent RF channels, each one equipped with a cryogenic low-noise amplifier. As KIDs are sensitive to magnetic fields, two magnetic shields were added to reduce this noise source: a mu-metal enclosure at 300 K and a superconducting lead screen on the 1~K stage. 
The experimental tool is optimised to work under low optical background representative of the in-space sky emission at 100~GHz and 150~GHz. The desired optical background was obtained using a testing device called {\it Sky Simulator} (SS). This device was originally built to mimic the typical optical background for the NIKA instrument at the IRAM 30~m telescope in Pico Veleta and rescaling it by regulating the diaphragm of the 100~mK lyot stop~\citep{catalano_3mm}. The spectral response of the detectors is measured with a Martin-Puplett interferometer~\citep{durand}. The pixels are back-illuminated through the silicon wafer.  We use a back-short cavity situated at an optimised distance, of 750~$\mu$m for the 3~mm array and
600~$\mu$m for the 2~mm array, to maximise the absorption of photons (Fig.\ref{fig:setup_1} left panel). For the CR impact characterisation measurements, we drilled the back-short in the centre as shown in Fig.\ref{fig:setup_1} right panel. We summarise the main characteristics of the experimental set-up  in Tab~\ref{tab:1}.

\begin{figure*}
\begin{center}
\includegraphics[width=8cm, , keepaspectratio]{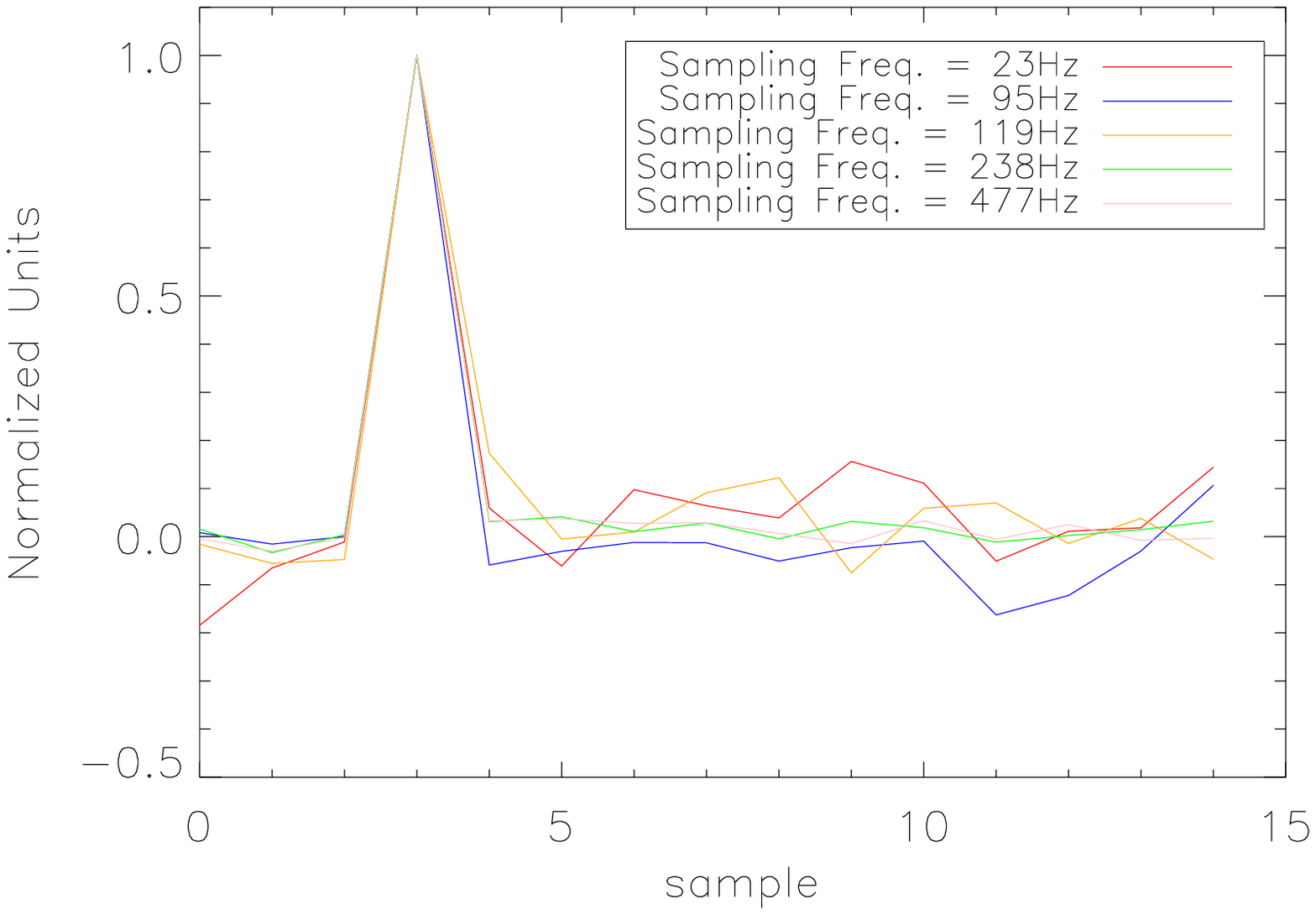}
\includegraphics[width=8cm, , keepaspectratio]{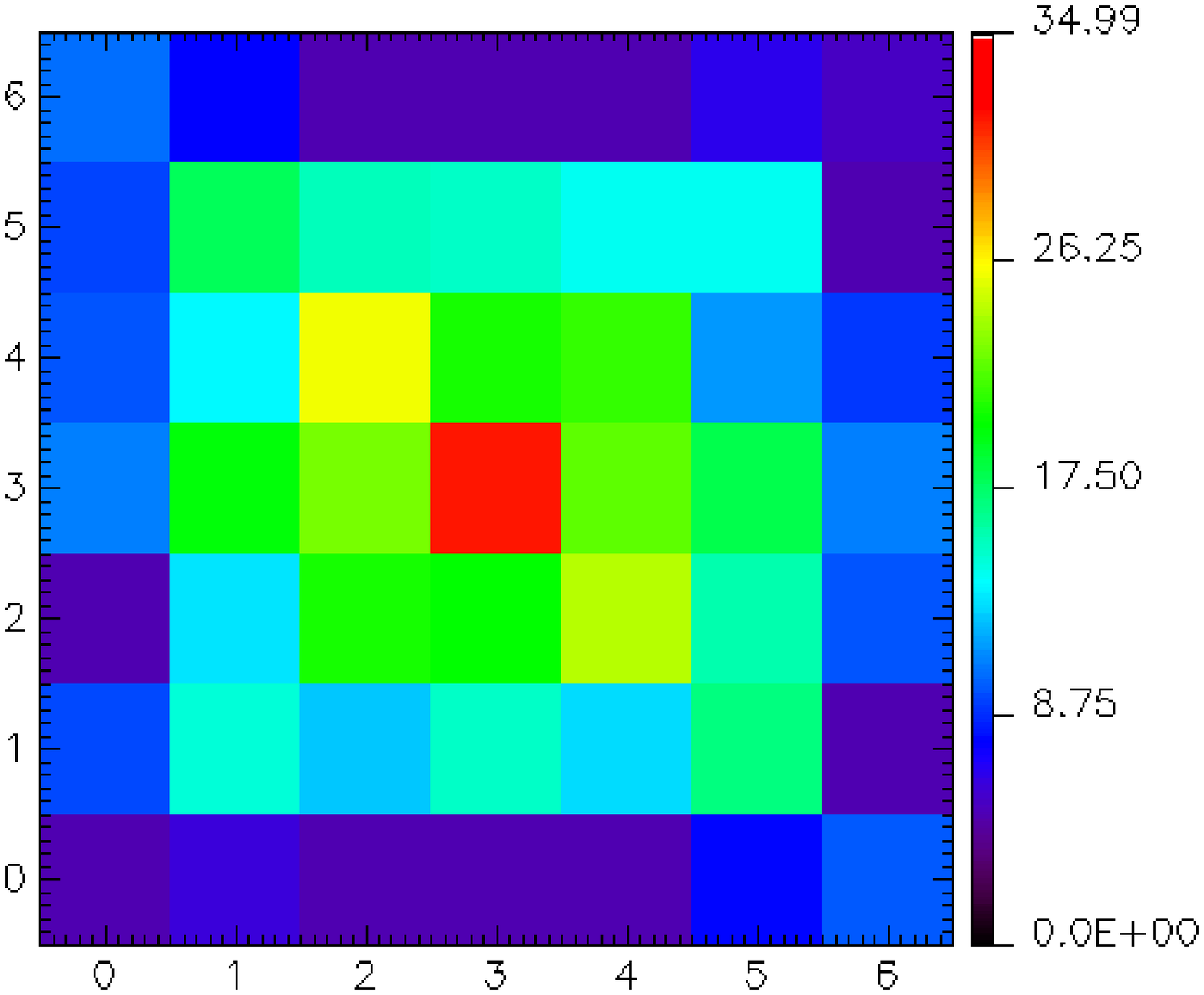}
\end{center}
\caption{Glitch characterisation results. Left panel: glitch template as a function of a sample for different read-out sampling frequency. Right panel: propagation of the energy in the LEKID array in units of signal-to-noise ratio. Each pixel in the map corresponds to a LEKID detector.}
\label{fig:glitch_kid}
\end{figure*}

\subsection{Optical response and Noise equivalent power}

The optical responsivity, proportional to the frequency shift of each resonance,  was measured using a vector network analyzer (VNA). The spectral response and noise was characterised with  NIKEL electronics~\citep{bourrion}, which was successfully used during several NIKA observing campaigns.

We perform frequency sweeps to measure the LEKID arrays transfer function for various SS background temperatures from 80 to 300~K. The frequency shift averaged across all the pixels correspond to about 27~kHz and 50~kHz for the 3~mm and 2~mm arrays, respectively, with a dispersion between detectors of about 30~\%. 
We performed an optical simulation of the system accounting
for the absorption, reflection, and emission of the polyethylene
lenses and the diffracted beam, which that is due to the cold aperture
stop at 100 mK, to estimate the optical background on the focal
plane~\citep{catalano_3mm}. The results of this simulation were validated by comparing
the optical background on NIKA Al arrays for laboratory
tests to that measured at the 30 m IRAM telescope. Considering the complexity of the set-up, however, we estimate the level of uncertainties to be about 50~\%. If we set the size of the cold aperture equal to 20~mm (resulting in nine times less optical background on the pixels then original condition) and we change the SS temperature from 40~K to 300~K, we can calculate the corresponding variation in optical power per pixel to about 0.6~pW for the 3~mm array and 1.8~pW for the 2~mm array. 

The spectral noise density, Sn(f) (in Hz/$\sqrt{Hz}$), is calculated at a fixed SS temperature of 80~K via NIKEL electronics. Correlated electronic noise is removed by subtracting a common mode. This is obtained by averaging the time-ordered-data (TOD) of all of the detectors in the array. The resulting template is fitted linearly to the TOD of each detector. The best fit is then subtracted from the detector TODs. After de-correlation, the spectral noise density is flat in a band between 1 and 10~Hz and equal to about $Sn(f) = 0.8-2$~Hz/$\sqrt{Hz}$ for the two arrays. We can compute the NEP as

\begin{equation}
NEP = \frac{Sn(f)}{\Re}
.\end{equation}

The left panel of Fig~\ref{fig:nep_kid} shows the distribution of the NEP for the 3~mm (red) and 2~mm (blue) arrays. 
The estimation of the required NEP depends on the particular working conditions of the instrument (e.g. spectral bandwidth and pixel size with respect to the telescope size). For the goals of this paper, we compare our results to a reference goal that has been defined as  $NEP_{GOAL} \leq 2 \cdot NEP_{phot}$, where $NEP_{phot}$ is photon noise that comes from the fluctuations of the incident radiation, as well as the goal chosen for the HFI detectors (vertical red and blue dashed lines in Fig~\ref{fig:nep_kid}). The right panel of Fig~\ref{fig:nep_kid} shows the corresponding normalised spectral response of the two arrays. The averaged NEP over the entire array is about twice the goal for both arrays and the best pixels approach the goals by few tens of percent.

\subsection{Glitch characterisation from ground measurements}\label{meas}

We used the same 2~mm LEKID array kept under the same optical background conditions for this test. We add an americium alpha particle source ($^{241}$Am) at a distance
of about 600~$\mu$m. The alpha particles hit the array on the front side (see Fig~\ref{fig:setup_1}, right panel). The $^{241}$Am source produces 5.4 MeV alpha particles, which are absorbed completely in the 300~$\mu$m silicon wafer. To rescale the absorbed energy to that corresponding to the pic of the in-space CR spectrum, we set a 10~$\mu$m copper shield in front of the source. This allows the reduction of Americium alpha particle energy  to normal distribution centred at 630~keV with a 30~keV 1 $\sigma$ dispersion (resulting from a Geant-4 simulation). This energy corresponds to the energy absorbed in the silicon wafer by a 200~MeV proton, which is the particle most typical of CRs at the second Lagrangian point. We performed a read-out of the LEKID array with the NIKEL electronics tuning the sampling frequency between 20 to 500~Hz. 
The main results of the test are described below.

\begin{table}[b!]
\begin{center}
\begin{tabular}{cc}
\hline
\hline
Sampling frequency [Hz] & Calibration factor  [kHz/keV]  \\
\hline \hline
23 &  20 \\
95 &  85  \\
119 & 100  \\
238 & 200  \\
477 & 420 \\
\hline \hline
\end{tabular}
\end{center}
\caption{Calibration factor derived from measurement for different read-out sampling frequencies.}
\label{tab:2}
\end{table}

\begin{itemize}

\item {\bf Suppression Factor:} As expected, the observed glitches are mostly constant in amplitude. This is because of the slightly fixed point-of-contact and impinging energy. Starting from the NEP measured in the previous section and the LEKID time constants measured in Sec~\ref{phy_int}, we derived calibration factors (in units kHz/keV) as a function of the sampling frequency. Results are presented in Tab~\ref{tab:2}. This allows us to derive a suppression factor $\chi$, which is defined as the ratio between the deposited energy and actual energy detected by the LEKID. This quantity is calculated as

$$
\chi = \frac{E_{\alpha} \cdot C}{Amp},
$$

\begin{figure*}
\begin{center}
\includegraphics[width=8cm, , keepaspectratio]{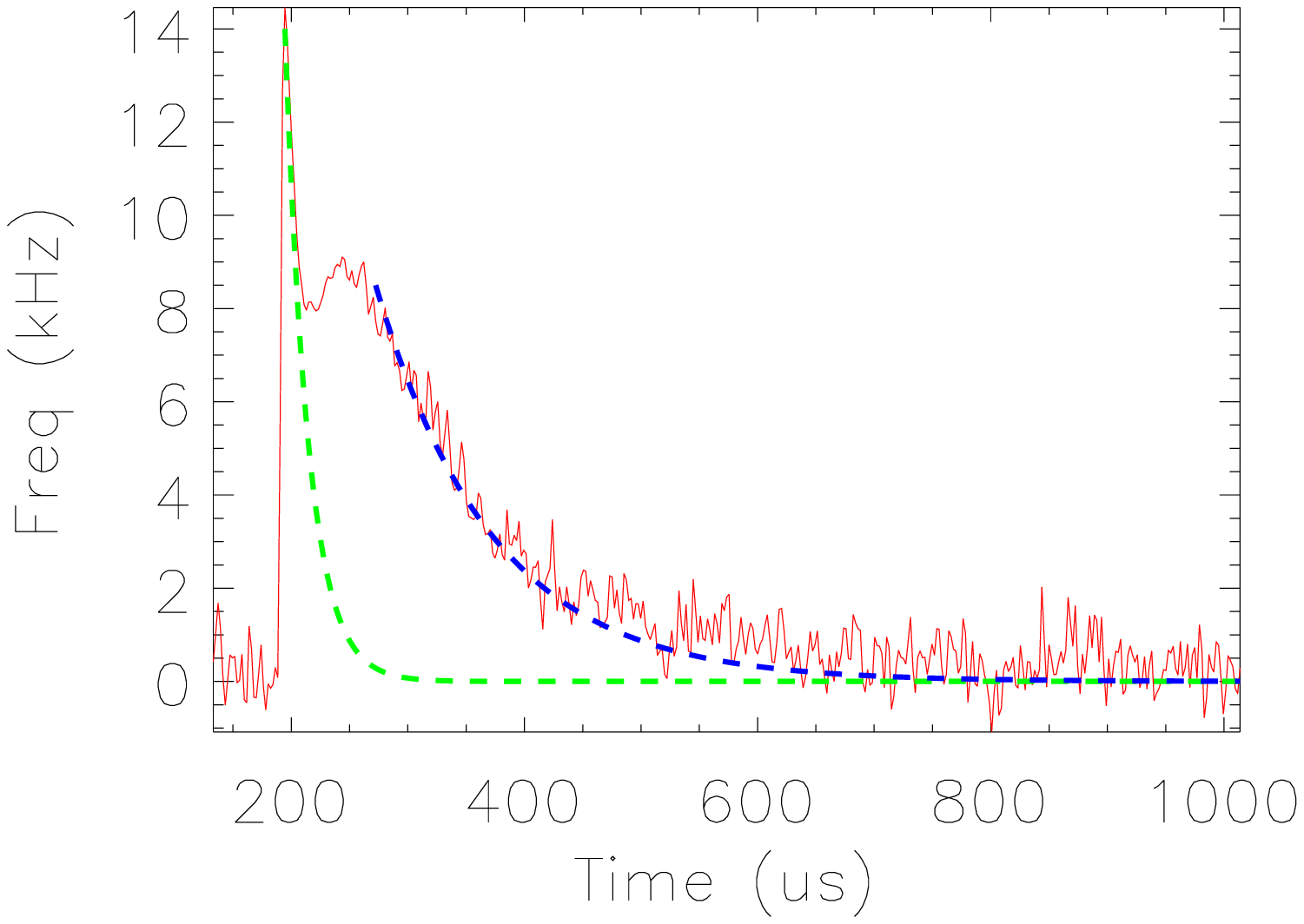}
\includegraphics[width=8cm, , keepaspectratio]{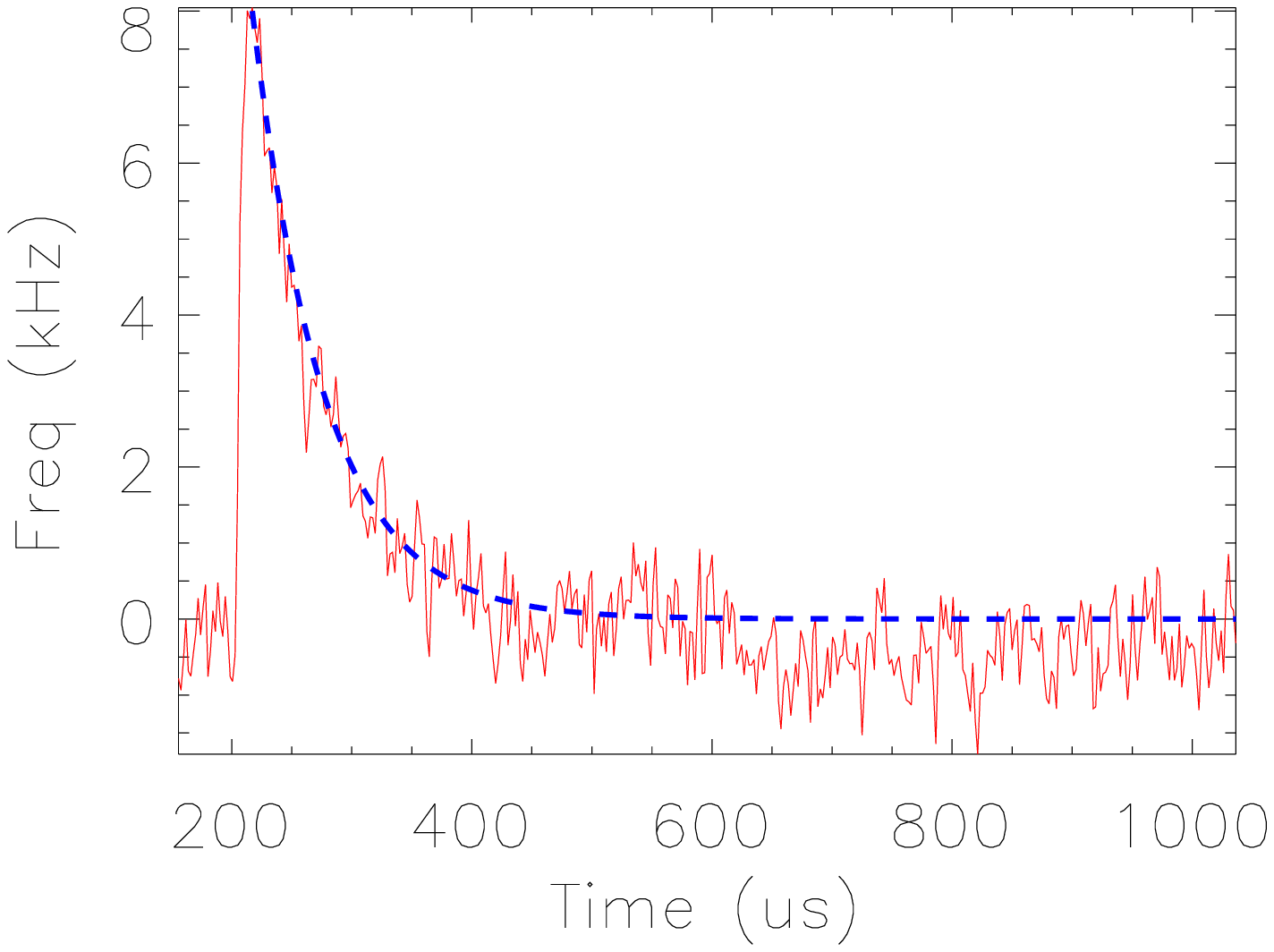}
\end{center}
\caption{Typical glitches produced from the hit of a Cd X-ray on the LEKID array, along with the time constants of a two exponential model (left panel) and one exponential model.}
\label{fig:nixa}
\end{figure*}

where $Amp$ is the maximum amplitude detected for each pixel measured in Hz, $C$ is the calibration factor, and $E_{\alpha}$ is the deposited energy of the $\alpha$ particle (630~keV).
We calculated the suppression factor at around 2000 and this value is nearly constant (as expected) for different read-out sampling frequencies. We estimated the variability of the suppression factor through the different pixels to be equal to 35\%. This means that when the particle hits the silicon very close to a LEKID, only 0.05~\% of its energy is transferred to the detector. The derived suppression factor is about three times larger then that measured between the silicon wafer and the high-impedance bolometers of HFI. 

\item {\bf Time Constants:} As shown in the left panel of Fig~\ref{fig:glitch_kid}, the glitch time constants
are unresolved for a range of sampling frequencies up to 500 Hz. This means that, in any case, for typical in-space read-out electronics, all glitches appear as one sample in the time ordered data instead of few tens of samples in the case of HFI bolometers. \cite{addabbo} indicated, although not fully confirmed, the presence of a few ms time constants. However, the measurements in this paper do not confirm this result and this kind of a slower component is not observed even when stacking few thousands of glitches.

\item {\bf Coincidences:} In terms of detected glitches at a level of 5~$\sigma$, the surface of the silicon wafer impacted by a 630~keV alpha particle never exceeds a square of 6x6 detectors (about 1.4~cm$^2$), as shown in Fig~\ref{fig:glitch_kid} (right panel). 

\end{itemize}

\subsubsection{Physical interpretation}\label{phy_int}
In parallel with the tests performed in typical space conditions (e.g. sampling rate and background), we performed measurements dedicated to characterising
the interaction of particles with LEKID detectors arrays using faster read-out electronics~\citep{bourrion_nixa}. This version of the electronics can acquire data fast enough to properly interpret the physical processes~\citep{addabbo}. The sampling rate can be tuned from 500~kHz to 2~MHz for a maximum of 12 channels over 500~MHz bandwidth.
\\
For these tests we used a Cadmium source that produces 25~keV X-rays and can impact all the detectors of the array with the same probability. If we analyse the observed glitches per detectors, two families of events can be isolated.  The first family (Fig \ref{fig:nixa}, left panel) peaks at an amplitude of between 12 and 14~kHz and corresponds to about 60~\% of the glitches. This family can be represented by a  double time-constant model (i.e. the faster between 10-15~$\mu$s and the slower between 80-100~$\mu$s).
The second family of glitches peaks at a lower amplitude (less then 10~kHz) and contains about 40\% of the glitches. These glitches agree reasonably well with a single time-constant model with the same slower 80-100~$\mu$s time constant as the first glitch family.
\\
A possible interpretation of these two time constants is presented below.  

\begin{itemize}

\item {\bf Fast time constant:} the dynamic response of a LEKID is determined, among other things, by the quasi-particle lifetime\footnote{The equilibrium state of a superconductor at a given temperature is represented by Cooper pair condensate and thermally excited quasiparticles.}~\citep{2008PhRvL.100y7002B}. This represents the time occurring between quasi-particle creation and their recombination into Cooper pairs following an excitation (phonon or photon) exceeding the superconducting gap. This time constant varies as a function of local quasi-particle density, meaning that we expect to have faster time constants for higher working temperature of the device or for stronger signal. Moore et al.~\citep{moore2012} have shown that this time constant is equal to about tens of microseconds for low-temperature aluminium film KID fabricated on a silicon substrate. The measured fast time constant of the first family of glitches is compatible with this process.

\item {\bf Slow time constant:} as we observed for spider-web bolometers, particles hitting the silicon wafer produce ballistic phonons that can propagate unhindered through the crystal over large distances (up to centimetres~\citep{holmes1999}). Typically, ballistic phonons decay within hundreds of microseconds into thermal phonons that cannot be sensed by LEKID detectors because their energy is lower than the superconducting gap. The slower time constant observed might be ascribed to this process.    

\end{itemize}   

Starting from this interpretation, we can explain the observed two families of events with the X-ray distance of impact with respect to the considered detector. Rescaling the results of the previous section, at this energy (25~keV) and at this read-out rate, an X-ray hitting the silicon wafer should produce a measurable signal (5~$\sigma$) on about nine detectors. When the X-ray impacts the silicon wafer close to the considered detectors, the time constant related to quasi-particle lifetime dominates in amplitude and therefore can be resolved; otherwise only the time constant related to the ballistic phonons propagation can be isolated.


\section{Simulation of the impact of cosmic rays in a LEKID array placed at L2}\label{sim_toi}

Starting from the results of the previous section, we developed an analytical model to rescale the results to what we would obtain if such a LEKID array would be operated at the second Lagrangian point. Using this model we can simulate a realistic timestream. In the following subsections, we describe the model and the generated timestream. 

\subsection{Toy proton model}

\begin{figure}[t!]
\begin{center}
\includegraphics[width=8cm, , keepaspectratio]{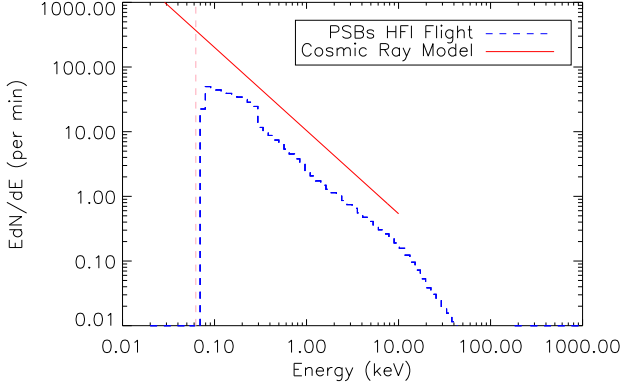}
\end{center}
\caption{LEKID toy proton model (red line) compared to a typical 143~GHz in-flight HFI bolometer (blue dashed line). The pink dashed line represents the LEKID 5~$\sigma$ level.}
\label{fig:sim1}
\end{figure}

We consider a solid square box made of silicon with the same 
volume of the tested LEKID arrays. We also consider that each detector has a surface equal to 2x2~mm$^2$. The inputs of the model are:

\begin{itemize}

\item geometrical parameters of the LEKID array;
\item stopping power function and density of the silicon;
\item energy distribution of CRs at the second Lagrangian point;

\item ratio between the absorbed energy in the substrate and the energy detected by the LEKID (suppression factor $\chi$). 

\end{itemize}

From the literature, we know the energy distribution of the proton at L2: 

\begin{equation}\label{cr_1eq}
\frac{\Delta N}{\Delta E_{p^+}} \sim E_{p^+}^{- \beta}
,\end{equation}

where $\beta $ is equal to 0.8. In the energy range of interest we can fit the stopping power function as a power law considering an impact angle equal to $\theta$. We obtain therefore the generic absorbed energy as a function of the proton energy and the impact angle

$$
E_{abs}=\frac{E_{O}}{cos \theta} \cdot (\frac{E_{p^+}}{E_{p^{+o}}})^{- \gamma}.
$$

Where $E_{p^+o}$ is the reference proton energy, $E_{O} = \rho_{sil} \cdot d \cdot SP(E_{p^+o})$ is the reference absorbed energy for an orthogonal impact with $\rho_{sil}$ the density of the silicon, $d$ is the thickness of the silicon die, and $SP(E_{p^+o})$ is the stopping power function calculated at a reference proton energy.
\\
By integrating Eq.~\ref{cr_1eq} over the solid angle, surface, and  integration time and considering the suppression factor, we obtain the energy absorbed in the detector,

\begin{equation}\label{cr_model}
\frac{\Delta N}{\Delta E_{LEKID}} = \frac{4 \pi A \Delta t E_{p^+o}\cdot \chi}{(2\gamma+\beta-1) \cdot E_{O}^{\frac{\beta-1}{\gamma}}} \cdot E_{abs}^{- \frac{\gamma + 1-\beta}{\gamma}}
,\end{equation}where $A$ is the surface of the silicon wafer impacted by a CR, $\Delta t$ is the integration time, and $\chi$ is the suppression factor. In Fig~\ref{fig:sim1} we show the spectrum obtained from Eq.~\ref{cr_model} compared to a glitch spectrum measured from a typical 143~GHz in-flight HFI bolometer.

\begin{figure}[b!]
\begin{center}
\includegraphics[width=9cm, , keepaspectratio]{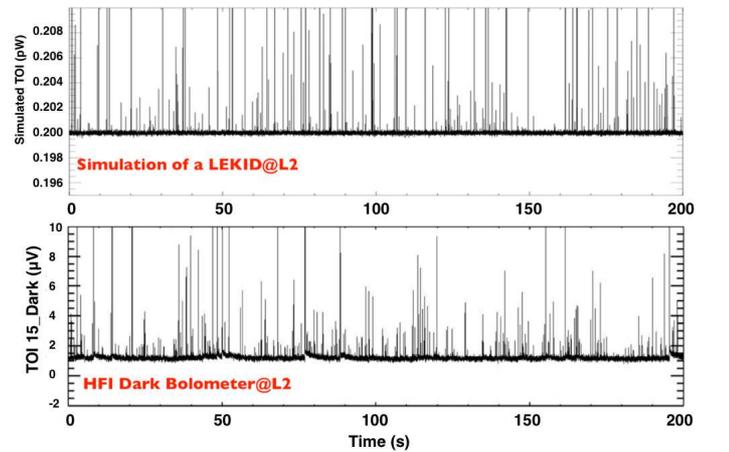}\\
\end{center}
\caption{Simulated timestream of a LEKID detector at L2 compared to an in-flight HFI dark bolometer.}
\label{fig:sim2}
\end{figure}

\begin{figure*}
\begin{center}
\includegraphics[width=8cm, , keepaspectratio]{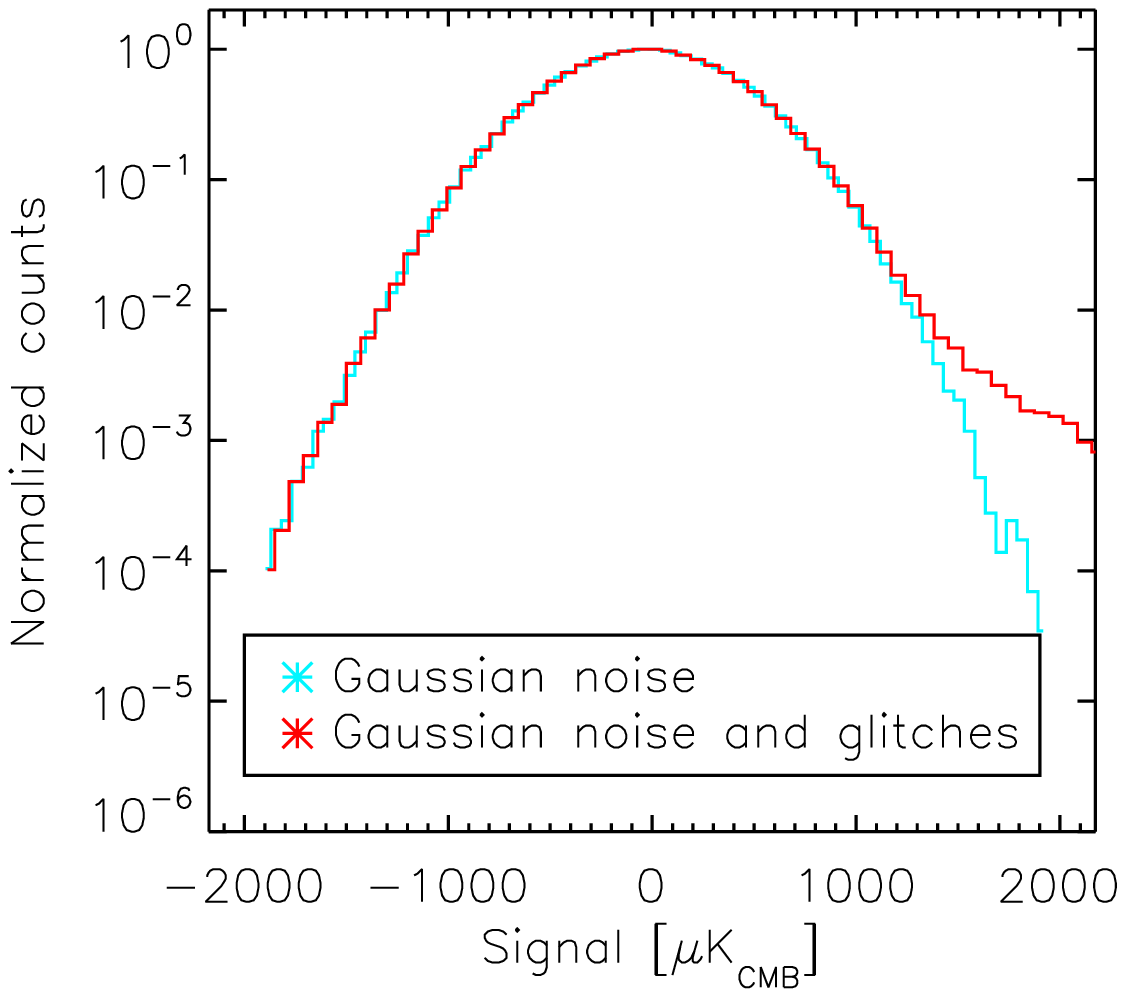}
\includegraphics[width=8cm, , keepaspectratio]{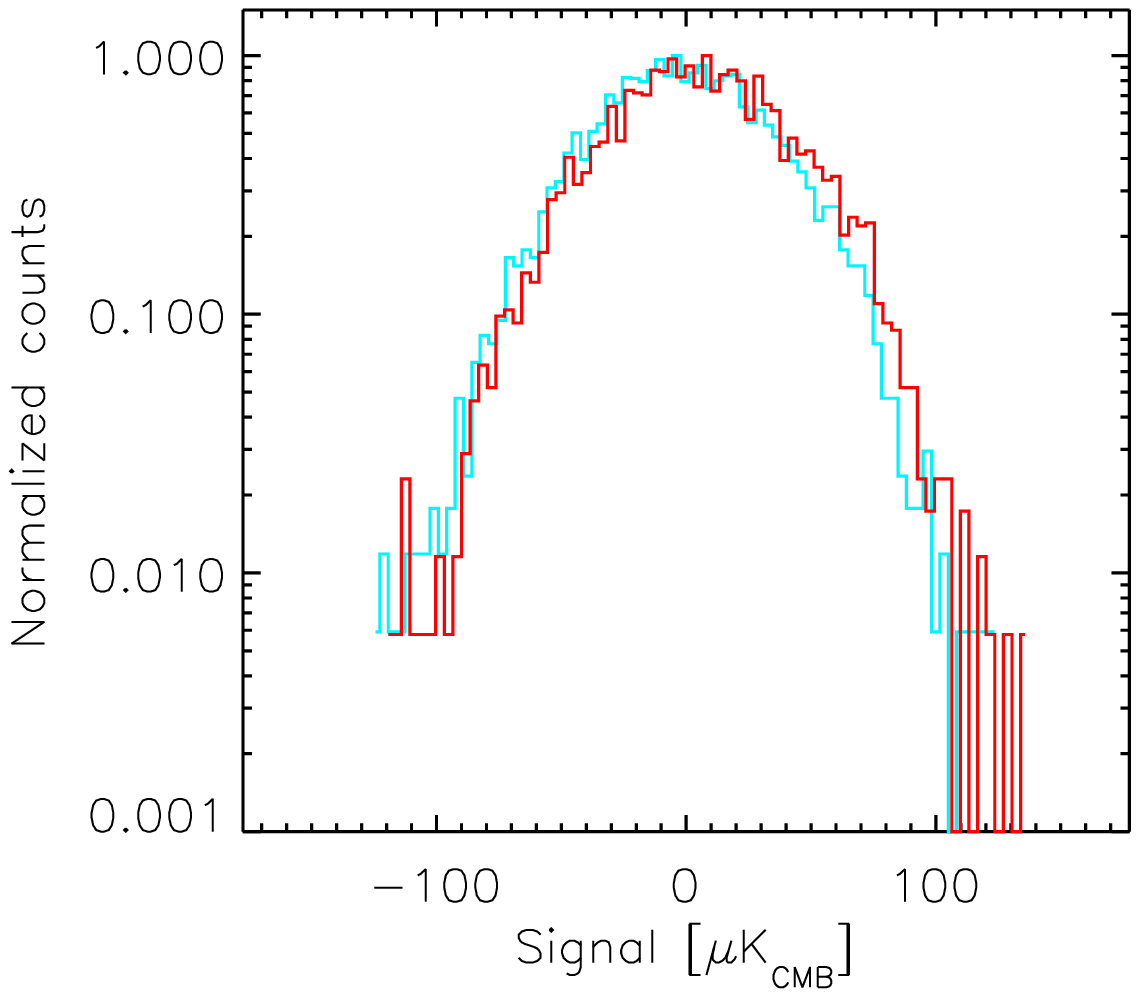}
\end{center}
\caption{Left panel: 1D distribution for simulated LEKID timestream with (red) and without (blue) glitch contribution. Right panel:  2D distribution for simulated LEKID sky map with (red) and without (blue) glitch contribution.}
\label{fig:sim3}
\end{figure*}

\subsection{Simulation of LEKID time-ordered data}

Starting from the distribution of the glitches absorbed energy induced by CRs, we simulated timestreams of LEKID data in the absence of sky signals. We added CR events to a realisation of noise with a standard deviation $\sigma$ given from the expected photon noise level with an average of 0.3~pW.
As already shown from laboratory measurements, the LEKID detectors have an unresolved time constant, so we consider that each glitch affects only one point of data.
The main results we can derive from this simulation are:

\begin{itemize}

\item {\bf Data loss:} As shown Fig~\ref{fig:sim2}, the CRs generate a rate of glitches of about 1.8~Hz, which is larger than what was observed in-flight on a typical Planck HFI bolometer. This is because, as described in Sec~\ref{meas}, the affected surface of the silicon wafer following a CR impact is about 1.4~cm$^2$, which is about twice that of the HFI bolometer silicon wafer. On the other hand, the time constants of the glitches are unresolved for sampling rates up to 500~Hz. The percent level of the flagged data due to the de-glitching is therefore about 1~\% compared to about 12-15~\% for Planck HFI bolometers.   

\item {\bf Glitch residual contamination:}
One of the main differences in comparing the HFI bolometer glitch impact and the LEKID array glitch impact is that for LEKID arrays, all detectors share the same substrate, while HFI bolometers are fully independent. As a consequence, CR impacts affect a surface of about 1.4~cm$^2$ in LEKID arrays giving a larger number of glitches per detectors. All of the residual glitches (below 5~$\sigma$) increase the rms noise by a factor of about 3\%. In terms of non-Gaussianity, we expect glitches to induce non-Gaussian features in the TOD. This is shown in the left panel of Fig~\ref{fig:sim3}, where we present the one-dimensional (1D) distribution for simulated TODs with (red) and without (blue) glitch contribution, assuming Gaussian detector noise at the photon noise level. Glitches show up as a positive tail with skewness and kurtosis more than 44~$\sigma$ away from that expected for the Gaussian noise. These non-Gaussian features observed in the TODs are significantly reduced when they are projected to construct maps of the sky. This is because the noise per pixel in the map is expected to decrease with the square root of the number of TOI samples per pixel, $\sqrt(N_{\mathrm{hits}}$, while the glitch contribution decreases with $N_{\rm hits}$. We project the TODs simulated above onto sky maps to test this
hypothesis. We assume a 2.5 year space mission and a typical sky pixel map size of 1 arcmin, which corresponds to Nhits = 150 samples per pixel for a Planck-like sampling rate.  We find that the glitch contribution leads to non-significant, non-Gaussianity in the final maps, as  the skewness and the kurtosis is less than 0.23~$\sigma$ away from that expected for the Gaussian noise. This can be observed in the right panel of Figure~\ref{fig:sim3} where we present the 1D distribution of the pixel values for the projected sky maps in the case of noise only (blue) and noise and glitches (red) simulations.

\end{itemize}

\section{Conclusion}

We have shown that LEKID arrays originally designed for ground-based measurements behave adequately in space-like conditions. Under space typical background conditions, the LEKID\ arrays show NEPs that are close to the CMB photon noise. Furthermore, although these arrays show a CR impact rate that is larger than HFI detectors, they present extremely short glitch time constants (not resolved up to 500~Hz sampling frequency) compared to the HFI bolometers time constants (from 5-10~ms up to 2~s). This makes the percent level of the flagged data due to the de-glitching of about 1 \% compared to about 12-15~\% for Planck HFI bolometers.
\\
Residual glitches (below 5~$\sigma$) add no significant contributions to the noise rms (less than 3~\%) and have negligible impact on non-Gaussianity studies.      

Starting from these promising results we think that the sensitivity could be improved by optimising, principally the resonator coupling to the RF feed-line, the meander geometry, and superconducting film thickness. 

In addition, the contribution from correlated electronic noise needs to be investigated in more detail to evaluate possible residuals after using standard map-making algorithms and/or decorrelation techniques for CMB science.

\begin{acknowledgement}
The engineers most involved in the experimental set-up development are Gregory Garde, Henri Rodenas, Jean-Paul Leggeri, Maurice Grollier, Guillaume Bres, Christophe Vescovi, Jean-Pierre Scordilis, and Eric Perbet. We acknowledge,  in general, the crucial contributions of the whole Cryogenics and Electronics groups at Institut N\'eel and LPSC. The arrays described in this paper have been produced at the CEA Saclay and PTA Grenoble microfabrication facilities. This work was supported as part of a collaborative project, SPACEKIDS, funded via grant 313320 provided by the European Commission under Theme SPA.2012.2.2-01 of Framework Programme 7.
\end{acknowledgement}

\bibliographystyle{aa}
\bibliography{cata}

\begin{thebibliography}{42}
\expandafter\ifx\csname natexlab\endcsname\relax\def\natexlab#1{#1}\fi

\bibitem[{{Adam} {et~al.}(2014){Adam}, {Comis}, {Mac{\'{\i}}as-P{\'e}rez},
  {Adane}, {Ade}, {Andr{\'e}}, {Beelen}, {Belier}, {Beno{\^i}t}, {Bideaud},
  {Billot}, {Boudou}, {Bourrion}, {Calvo}, {Catalano}, {Coiffard}, {D'Addabbo},
  {D{\'e}sert}, {Doyle}, {Goupy}, {Kramer}, {Leclercq}, {Martino}, {Mauskopf},
  {Mayet}, {Monfardini}, {Pajot}, {Pascale}, {Perotto}, {Pointecouteau},
  {Ponthieu}, {Rev{\'e}ret}, {Rodriguez}, {Savini}, {Schuster}, {Sievers},
  {Tucker}, \& {Zylka}}]{adam2014}
{Adam}, R., {Comis}, B., {Mac{\'{\i}}as-P{\'e}rez}, J.~F., {et~al.} 2014, \aap,
  569, A66

\bibitem[{{Ade} {et~al.}(2014){Ade}, {Aikin}, {Amiri}, {Barkats}, {Benton},
  {Bischoff}, {Bock}, {Brevik}, {Buder}, {Bullock}, {Davis}, {Day}, {Dowell},
  {Duband}, {Filippini}, {Fliescher}, {Golwala}, {Halpern}, {Hasselfield},
  {Hildebrandt}, {Hilton}, {Irwin}, {Karkare}, {Kaufman}, {Keating},
  {Kernasovskiy}, {Kovac}, {Kuo}, {Leitch}, {Llombart}, {Lueker},
  {Netterfield}, {Nguyen}, {O'Brient}, {Ogburn}, {Orlando}, {Pryke},
  {Reintsema}, {Richter}, {Schwarz}, {Sheehy}, {Staniszewski}, {Story},
  {Sudiwala}, {Teply}, {Tolan}, {Turner}, {Vieregg}, {Wilson}, {Wong}, {Yoon},
  \& {Bicep2 Collaboration}}]{2014ApJ...792...62A}
{Ade}, P.~A.~R., {Aikin}, R.~W., {Amiri}, M., {et~al.} 2014, \apj, 792, 62

\bibitem[{{Andr{\'e}} {et~al.}(2014){Andr{\'e}}, {Baccigalupi}, {Banday},
  {Barbosa}, {Barreiro}, {Bartlett}, {Bartolo}, {Battistelli}, {Battye},
  {Bendo}, {Beno{\^i}t}, {Bernard}, {Bersanelli}, {B{\'e}thermin}, {Bielewicz},
  {Bonaldi}, {Bouchet}, {Boulanger}, {Brand}, {Bucher}, {Burigana}, {Cai},
  {Camus}, {Casas}, {Casasola}, {Castex}, {Challinor}, {Chluba}, {Chon},
  {Colafrancesco}, {Comis}, {Cuttaia}, {D'Alessandro}, {Da Silva}, {Davis}, {de
  Avillez}, {de Bernardis}, {de Petris}, {de Rosa}, {de Zotti}, {Delabrouille},
  {D{\'e}sert}, {Dickinson}, {Diego}, {Dunkley}, {En{\ss}lin}, {Errard},
  {Falgarone}, {Ferreira}, {Ferri{\`e}re}, {Finelli}, {Fletcher}, {Fosalba},
  {Fuller}, {Galli}, {Ganga}, {Garc{\'{\i}}a-Bellido}, {Ghribi}, {Giard},
  {Giraud-H{\'e}raud}, {Gonzalez-Nuevo}, {Grainge}, {Gruppuso}, {Hall},
  {Hamilton}, {Haverkorn}, {Hernandez-Monteagudo}, {Herranz}, {Jackson},
  {Jaffe}, {Khatri}, {Kunz}, {Lamagna}, {Lattanzi}, {Leahy}, {Lesgourgues},
  {Liguori}, {Liuzzo}, {Lopez-Caniego}, {Macias-Perez}, {Maffei}, {Maino},
  {Mangilli}, {Martinez-Gonzalez}, {Martins}, {Masi}, {Massardi}, {Matarrese},
  {Melchiorri}, {Melin}, {Mennella}, {Mignano}, {Miville-Desch{\^e}nes},
  {Monfardini}, {Murphy}, {Naselsky}, {Nati}, {Natoli}, {Negrello}, {Noviello},
  {O'Sullivan}, {Paci}, {Pagano}, {Paladino}, {Palanque-Delabrouille},
  {Paoletti}, {Peiris}, {Perrotta}, {Piacentini}, {Piat}, {Piccirillo},
  {Pisano}, {Polenta}, {Pollo}, {Ponthieu}, {Remazeilles}, {Ricciardi},
  {Roman}, {Rosset}, {Rubino-Martin}, {Salatino}, {Schillaci}, {Shellard},
  {Silk}, {Starobinsky}, {Stompor}, {Sunyaev}, {Tartari}, {Terenzi},
  {Toffolatti}, {Tomasi}, {Trappe}, {Tristram}, {Trombetti}, {Tucci}, {Van de
  Weijgaert}, {Van Tent}, {Verde}, {Vielva}, {Wandelt}, {Watson}, \&
  {Withington}}]{2014JCAP...02..006A}
{Andr{\'e}}, P., {Baccigalupi}, C., {Banday}, A., {et~al.} 2014, \jcap, 2, 006

\bibitem[{{Barends} {et~al.}(2008){Barends}, {Baselmans}, {Yates}, {Gao},
  {Hovenier}, \& {Klapwijk}}]{2008PhRvL.100y7002B}
{Barends}, R., {Baselmans}, J.~J.~A., {Yates}, S.~J.~C., {et~al.} 2008,
  Physical Review Letters, 100, 257002

\bibitem[{{Bennett} {et~al.}(2013){Bennett}, {Larson}, {Weiland}, {Jarosik},
  {Hinshaw}, {Odegard}, {Smith}, {Hill}, {Gold}, {Halpern}, {Komatsu}, {Nolta},
  {Page}, {Spergel}, {Wollack}, {Dunkley}, {Kogut}, {Limon}, {Meyer}, {Tucker},
  \& {Wright}}]{wmap}
{Bennett}, C.~L., {Larson}, D., {Weiland}, J.~L., {et~al.} 2013, \apjs, 208, 20

\bibitem[{{Beno{\^i}t} {et~al.}(2004){Beno{\^i}t}, {Ade}, {Amblard}, {Ansari},
  {Aubourg}, {Bargot}, {Bartlett}, {Bernard}, {Bhatia}, {Blanchard}, {Bock},
  {Boscaleri}, {Bouchet}, {Bourrachot}, {Camus}, {Couchot}, {de Bernardis},
  {Delabrouille}, {D{\'e}sert}, {Dor{\'e}}, {Douspis}, {Dumoulin}, {Dupac},
  {Filliatre}, {Fosalba}, {Ganga}, {Gannaway}, {Gautier}, {Giard},
  {Giraud-H{\'e}raud}, {Gispert}, {Guglielmi}, {Hamilton}, {Hanany},
  {Henrot-Versill{\'e}}, {Kaplan}, {Lagache}, {Lamarre}, {Lange},
  {Mac{\'{\i}}as-P{\'e}rez}, {Madet}, {Maffei}, {Magneville}, {Marrone},
  {Masi}, {Mayet}, {Murphy}, {Naraghi}, {Nati}, {Patanchon}, {Perrin}, {Piat},
  {Ponthieu}, {Prunet}, {Puget}, {Renault}, {Rosset}, {Santos}, {Starobinsky},
  {Strukov}, {Sudiwala}, {Teyssier}, {Tristram}, {Tucker}, {Vanel}, {Vibert},
  {Wakui}, \& {Yvon}}]{2004A&A...424..571B}
{Beno{\^i}t}, A., {Ade}, P., {Amblard}, A., {et~al.} 2004, \aap, 424, 571

\bibitem[{{Bourrion} {et~al.}(2012){Bourrion}, {Vescovi}, {Bouly}, {Benoit},
  {Calvo}, {Gallin-Martel}, {Macias-Perez}, \& {Monfardini}}]{bourrion}
{Bourrion}, O., {Vescovi}, C., {Bouly}, J.~L., {et~al.} 2012, in Society of
  Photo-Optical Instrumentation Engineers (SPIE) Conference Series, Vol. 8452,
  Society of Photo-Optical Instrumentation Engineers (SPIE) Conference Series

\bibitem[{{Bourrion} {et~al.}(2013){Bourrion}, {Vescovi}, {Catalano}, {Calvo},
  {D'Addabbo}, {Goupy}, {Boudou}, {Macias-Perez}, \&
  {Monfardini}}]{bourrion_nixa}
{Bourrion}, O., {Vescovi}, C., {Catalano}, A., {et~al.} 2013, Journal of
  Instrumentation, 8, C2006

\bibitem[{{Catalano} {et~al.}(2014{\natexlab{a}}){Catalano}, {Ade}, {Atik},
  {Benoit}, {Br{\'e}ele}, {Bock}, {Camus}, {Chabot}, {Charra}, {Crill},
  {Coron}, {Coulais}, {D{\'e}sert}, {Fauvet}, {Giraud-H{\'e}raud},
  {Guillaudin}, {Holmes}, {Jones}, {Lamarre}, {Mac{\'{\i}}as-P{\'e}rez},
  {Martinez}, {Miniussi}, {Monfardini}, {Pajot}, {Patanchon}, {Pelissier},
  {Piat}, {Puget}, {Renault}, {Rosset}, {Santos}, {Sauv{\'e}}, {Spencer}, \&
  {Sudiwala}}]{catalano_glitch}
{Catalano}, A., {Ade}, P., {Atik}, Y., {et~al.} 2014{\natexlab{a}}, \aap, 569,
  A88

\bibitem[{{Catalano} {et~al.}(2014{\natexlab{b}}){Catalano}, {Calvo},
  {Ponthieu}, {Adam}, {Adane}, {Ade}, {Andr{\'e}}, {Beelen}, {Belier},
  {Beno{\^i}t}, {Bideaud}, {Billot}, {Boudou}, {Bourrion}, {Coiffard}, {Comis},
  {D'Addabbo}, {D{\'e}sert}, {Doyle}, {Goupy}, {Kramer}, {Leclercq},
  {Mac{\'{\i}}as-P{\'e}rez}, {Martino}, {Mauskopf}, {Mayet}, {Monfardini},
  {Pajot}, {Pascale}, {Perotto}, {Rev{\'e}ret}, {Rodriguez}, {Savini},
  {Schuster}, {Sievers}, {Tucker}, \& {Zylka}}]{catalano_nika2014}
{Catalano}, A., {Calvo}, M., {Ponthieu}, N., {et~al.} 2014{\natexlab{b}}, \aap,
  569, A9

\bibitem[{{Catalano} {et~al.}(2015){Catalano}, {Goupy}, {le Sueur}, {Benoit},
  {Bourrion}, {Calvo}, {D'addabbo}, {Dumoulin}, {Levy-Bertrand},
  {Mac{\'{\i}}as-P{\'e}rez}, {Marnieros}, {Ponthieu}, \&
  {Monfardini}}]{catalano_3mm}
{Catalano}, A., {Goupy}, J., {le Sueur}, H., {et~al.} 2015, \aap, 580, A15

\bibitem[{{D'Addabbo} {et~al.}(2014){D'Addabbo}, {Calvo}, {Goupy}, {Benoit},
  {Bourrion}, {Catalano}, {Macias-Perez}, \& {Monfardini}}]{addabbo}
{D'Addabbo}, A., {Calvo}, M., {Goupy}, J., {et~al.} 2014, in Society of
  Photo-Optical Instrumentation Engineers (SPIE) Conference Series, Vol. 9153,
  Society of Photo-Optical Instrumentation Engineers (SPIE) Conference Series,
  2

\bibitem[{Doyle(2008)}]{doyle}
Doyle, S. 2008, PhD Thesis, 1, 193

\bibitem[{Durand(2007)}]{durand}
Durand, T. 2007, Theses, {Universit{\'e} Joseph-Fourier - Grenoble I}

\bibitem[{{Fixsen} {et~al.}(1996){Fixsen}, {Cheng}, {Gales}, {Mather},
  {Shafer}, \& {Wright}}]{Fixsen}
{Fixsen}, D.~J., {Cheng}, E.~S., {Gales}, J.~M., {et~al.} 1996, \apj, 473, 576

\bibitem[{{Griffin} {et~al.}(2010){Griffin}, {Abergel}, {Abreu}, {Ade},
  {Andr{\'e}}, {Augueres}, {Babbedge}, {Bae}, {Baillie}, {Baluteau}, {Barlow},
  {Bendo}, {Benielli}, {Bock}, {Bonhomme}, {Brisbin}, {Brockley-Blatt},
  {Caldwell}, {Cara}, {Castro-Rodriguez}, {Cerulli}, {Chanial}, {Chen},
  {Clark}, {Clements}, {Clerc}, {Coker}, {Communal}, {Conversi}, {Cox},
  {Crumb}, {Cunningham}, {Daly}, {Davis}, {de Antoni}, {Delderfield}, {Devin},
  {di Giorgio}, {Didschuns}, {Dohlen}, {Donati}, {Dowell}, {Dowell}, {Duband},
  {Dumaye}, {Emery}, {Ferlet}, {Ferrand}, {Fontignie}, {Fox}, {Franceschini},
  {Frerking}, {Fulton}, {Garcia}, {Gastaud}, {Gear}, {Glenn}, {Goizel},
  {Griffin}, {Grundy}, {Guest}, {Guillemet}, {Hargrave}, {Harwit}, {Hastings},
  {Hatziminaoglou}, {Herman}, {Hinde}, {Hristov}, {Huang}, {Imhof}, {Isaak},
  {Israelsson}, {Ivison}, {Jennings}, {Kiernan}, {King}, {Lange}, {Latter},
  {Laurent}, {Laurent}, {Leeks}, {Lellouch}, {Levenson}, {Li}, {Li},
  {Lilienthal}, {Lim}, {Liu}, {Lu}, {Madden}, {Mainetti}, {Marliani}, {McKay},
  {Mercier}, {Molinari}, {Morris}, {Moseley}, {Mulder}, {Mur}, {Naylor},
  {Nguyen}, {O'Halloran}, {Oliver}, {Olofsson}, {Olofsson}, {Orfei}, {Page},
  {Pain}, {Panuzzo}, {Papageorgiou}, {Parks}, {Parr-Burman}, {Pearce},
  {Pearson}, {P{\'e}rez-Fournon}, {Pinsard}, {Pisano}, {Podosek}, {Pohlen},
  {Polehampton}, {Pouliquen}, {Rigopoulou}, {Rizzo}, {Roseboom}, {Roussel},
  {Rowan-Robinson}, {Rownd}, {Saraceno}, {Sauvage}, {Savage}, {Savini},
  {Sawyer}, {Scharmberg}, {Schmitt}, {Schneider}, {Schulz}, {Schwartz},
  {Shafer}, {Shupe}, {Sibthorpe}, {Sidher}, {Smith}, {Smith}, {Smith},
  {Spencer}, {Stobie}, {Sudiwala}, {Sukhatme}, {Surace}, {Stevens}, {Swinyard},
  {Trichas}, {Tourette}, {Triou}, {Tseng}, {Tucker}, {Turner}, {Vaccari},
  {Valtchanov}, {Vigroux}, {Virique}, {Voellmer}, {Walker}, {Ward}, {Waskett},
  {Weilert}, {Wesson}, {White}, {Whitehouse}, {Wilson}, {Winter}, {Woodcraft},
  {Wright}, {Xu}, {Zavagno}, {Zemcov}, {Zhang}, \& {Zonca}}]{Griffin}
{Griffin}, M.~J., {Abergel}, A., {Abreu}, A., {et~al.} 2010, \aap, 518, L3

\bibitem[{{Hlozek} \& {ACT Collaboration}(2013)}]{2013AAS...22110502H}
{Hlozek}, R. \& {ACT Collaboration}. 2013, in American Astronomical Society
  Meeting Abstracts, Vol. 221, American Astronomical Society Meeting Abstracts
  221, 105.02

\bibitem[{{Holmes} {et~al.}(1998){Holmes}, {Gildemeister}, {Richards}, \&
  {Kotsubo}}]{holmes1999}
{Holmes}, W., {Gildemeister}, J.~M., {Richards}, P.~L., \& {Kotsubo}, V. 1998,
  Applied Physics Letters, 72, 2250

\bibitem[{{Horeau} {et~al.}(2012){Horeau}, {Boulade}, {Claret}, {Feuchtgruber},
  {Okumura}, {Panuzzo}, {Papageorgiou}, {Rev{\'e}ret}, {Rodriguez}, \&
  {Sauvage}}]{2012arXiv1207.5597H}
{Horeau}, B., {Boulade}, O., {Claret}, A., {et~al.} 2012, ArXiv e-prints

\bibitem[{{Hu} \& {White}(1997)}]{HuWhite}
{Hu}, W. \& {White}, M. 1997, \na, 2, 323

\bibitem[{{Jaffe} {et~al.}(2003){Jaffe}, {Abroe}, {Borrill}, {Collins},
  {Ferreira}, {Hanany}, {Johnson}, {Lee}, {Matsumura}, {Rabii}, {Renbarger},
  {Richards}, {Smoot}, {Stompor}, {Tran}, {Winant}, \& {Proty
  Wu}}]{2003NewAR..47..727J}
{Jaffe}, A.~H., {Abroe}, M., {Borrill}, J., {et~al.} 2003, \nar, 47, 727

\bibitem[{{Kogut} {et~al.}(1996){Kogut}, {Banday}, {Bennett}, {Gorski},
  {Hinshaw}, {Jackson}, {Keegstra}, {Lineweaver}, {Smoot}, {Tenorio}, \&
  {Wright}}]{cobe}
{Kogut}, A., {Banday}, A.~J., {Bennett}, C.~L., {et~al.} 1996, \apj, 470, 653

\bibitem[{{Kogut} {et~al.}(2011){Kogut}, {Fixsen}, {Chuss}, {Dotson}, {Dwek},
  {Halpern}, {Hinshaw}, {Meyer}, {Moseley}, {Seiffert}, {Spergel}, \&
  {Wollack}}]{kogut2011}
{Kogut}, A., {Fixsen}, D.~J., {Chuss}, D.~T., {et~al.} 2011, \jcap, 7, 25

\bibitem[{{Kovac} {et~al.}(2002){Kovac}, {Leitch}, {Pryke}, {Carlstrom},
  {Halverson}, \& {Holzapfel}}]{2002Natur.420..772K}
{Kovac}, J.~M., {Leitch}, E.~M., {Pryke}, C., {et~al.} 2002, \nat, 420, 772

\bibitem[{{Leske} {et~al.}(2011){Leske}, {Cummings}, {Mewaldt}, \&
  {Stone}}]{Leske2011}
{Leske}, R.~A., {Cummings}, A.~C., {Mewaldt}, R.~A., \& {Stone}, E.~C. 2011,
  \ssr, 126

\bibitem[{{MacTavish} {et~al.}(2006){MacTavish}, {Ade}, {Bock}, {Bond},
  {Borrill}, {Boscaleri}, {Cabella}, {Contaldi}, {Crill}, {de Bernardis}, {De
  Gasperis}, {de Oliveira-Costa}, {De Troia}, {di Stefano}, {Hivon}, {Jaffe},
  {Jones}, {Kisner}, {Lange}, {Lewis}, {Masi}, {Mauskopf}, {Melchiorri},
  {Montroy}, {Natoli}, {Netterfield}, {Pascale}, {Piacentini}, {Pogosyan},
  {Polenta}, {Prunet}, {Ricciardi}, {Romeo}, {Ruhl}, {Santini}, {Tegmark},
  {Veneziani}, \& {Vittorio}}]{2006ApJ...647..799M}
{MacTavish}, C.~J., {Ade}, P.~A.~R., {Bock}, J.~J., {et~al.} 2006, \apj, 647,
  799

\bibitem[{{Matsumura} {et~al.}(2014){Matsumura}, {Akiba}, {Borrill}, {Chinone},
  {Dobbs}, {Fuke}, {Ghribi}, {Hasegawa}, {Hattori}, {Hattori}, {Hazumi},
  {Holzapfel}, {Inoue}, {Ishidoshiro}, {Ishino}, {Ishitsuka}, {Karatsu},
  {Katayama}, {Kawano}, {Kibayashi}, {Kibe}, {Kimura}, {Kimura}, {Koga},
  {Kozu}, {Komatsu}, {Lee}, {Matsuhara}, {Mima}, {Mitsuda}, {Mizukami},
  {Morii}, {Morishima}, {Murayama}, {Nagai}, {Nagata}, {Nakamura}, {Naruse},
  {Natsume}, {Nishibori}, {Nishino}, {Noda}, {Noguchi}, {Ogawa}, {Oguri},
  {Ohta}, {Otani}, {Richards}, {Sakai}, {Sato}, {Sato}, {Sekimoto}, {Shimizu},
  {Shinozaki}, {Sugita}, {Suzuki}, {Suzuki}, {Tajima}, {Takada}, {Takakura},
  {Takei}, {Tomaru}, {Uzawa}, {Wada}, {Watanabe}, {Yoshida}, {Yamasaki},
  {Yoshida}, \& {Yotsumoto}}]{hazumi}
{Matsumura}, T., {Akiba}, Y., {Borrill}, J., {et~al.} 2014, Journal of Low
  Temperature Physics, 176, 733

\bibitem[{{Mauskopf} {et~al.}(2014){Mauskopf}, {Doyle}, {Barry}, {Rowe},
  {Bidead}, {Ade}, {Tucker}, {Castillo}, {Monfardini}, {Goupy}, \&
  {Calvo}}]{Mauskopf}
{Mauskopf}, P., {Doyle}, S., {Barry}, P., {et~al.} 2014, Journal of Low
  Temperature Physics, 176, 545

\bibitem[{{Mewaldt} {et~al.}(2010){Mewaldt}, {Davis}, {Lave}, {Leske}, {Stone},
  {Wiedenbeck}, {Binns}, {Christian}, {Cummings}, {de Nolfo}, {Israel},
  {Labrador}, \& {von Rosenvinge}}]{Mewaldt2010}
{Mewaldt}, R.~A., {Davis}, A.~J., {Lave}, K.~A., {et~al.} 2010, \apjl, 723, L1

\bibitem[{{Monfardini} {et~al.}(2014){Monfardini}, {Adam}, {Adane}, {Ade},
  {Andr{\'e}}, {Beelen}, {Belier}, {Benoit}, {Bideaud}, {Billot}, {Bourrion},
  {Calvo}, {Catalano}, {Coiffard}, {Comis}, {D'Addabbo}, {D{\'e}sert}, {Doyle},
  {Goupy}, {Kramer}, {Leclercq}, {Macias-Perez}, {Martino}, {Mauskopf},
  {Mayet}, {Pajot}, {Pascale}, {Ponthieu}, {Rev{\'e}ret}, {Rodriguez},
  {Savini}, {Schuster}, {Sievers}, {Tucker}, \& {Zylka}}]{2014JLTP..176..787M}
{Monfardini}, A., {Adam}, R., {Adane}, A., {et~al.} 2014, Journal of Low
  Temperature Physics, 176, 787

\bibitem[{{Monfardini} {et~al.}(2011){Monfardini}, {Benoit}, {Bideaud},
  {Swenson}, {Cruciani}, {Camus}, {Hoffmann}, {D{\'e}sert}, {Doyle}, {Ade},
  {Mauskopf}, {Tucker}, {Roesch}, {Leclercq}, {Schuster}, {Endo}, {Baryshev},
  {Baselmans}, {Ferrari}, {Yates}, {Bourrion}, {Macias-Perez}, {Vescovi},
  {Calvo}, \& {Giordano}}]{monfardini2012}
{Monfardini}, A., {Benoit}, A., {Bideaud}, A., {et~al.} 2011, \apjs, 194, 24

\bibitem[{{Moore} {et~al.}(2012){Moore}, {Golwala}, {Bumble}, {Cornell}, {Day},
  {LeDuc}, \& {Zmuidzinas}}]{moore2012}
{Moore}, D.~C., {Golwala}, S.~R., {Bumble}, B., {et~al.} 2012, Applied Physics
  Letters, 100, 232601

\bibitem[{{Pilbratt} \& {Vandenbussche}(2012)}]{hershel}
{Pilbratt}, G. \& {Vandenbussche}, B. 2012, in COSPAR Meeting, Vol.~39, 39th
  COSPAR Scientific Assembly, 1499

\bibitem[{{Planck Collaboration} {et~al.}(2014{\natexlab{a}}){Planck
  Collaboration}, {Ade}, {Aghanim}, {Armitage-Caplan}, {Arnaud}, {Ashdown},
  {Atrio-Barandela}, {Aumont}, {Baccigalupi}, {Banday}, \&
  et~al.}]{Planck2013glitch}
{Planck Collaboration}, {Ade}, P.~A.~R., {Aghanim}, N., {et~al.}
  2014{\natexlab{a}}, \aap, 571, A10

\bibitem[{{Planck Collaboration} {et~al.}(2014{\natexlab{b}}){Planck
  Collaboration}, {Ade}, {Aghanim}, {Armitage-Caplan}, {Arnaud}, {Ashdown},
  {Atrio-Barandela}, {Aumont}, {Baccigalupi}, {Banday}, \& et~al.}]{cosmo}
{Planck Collaboration}, {Ade}, P.~A.~R., {Aghanim}, N., {et~al.}
  2014{\natexlab{b}}, \aap, 571, A16

\bibitem[{{Planck Collaboration (2013 results I)}(2014)}]{planck1}
{Planck Collaboration (2013 results I)}. 2014, \aap, in press,
  [arXiv:astro-ph/1303.5062]

\bibitem[{{Planck HFI Core Team} {et~al.}(2011){Planck HFI Core Team}, {Ade},
  {Aghanim}, {Ansari}, {Arnaud}, {Ashdown}, {Aumont}, {Banday}, {Bartelmann},
  {Bartlett}, {Battaner}, {Benabed}, {Beno{\^i}t}, {Bernard}, {Bersanelli},
  {Bhatia}, {Bock}, {Bond}, {Borrill}, {Bouchet}, {Boulanger}, {Bradshaw},
  {Br{\'e}elle}, {Bucher}, {Camus}, {Cardoso}, {Catalano}, {Challinor},
  {Chamballu}, {Charra}, {Charra}, {Chary}, {Chiang}, {Church}, {Clements},
  {Colombi}, {Couchot}, {Coulais}, {Cressiot}, {Crill}, {Crook}, {de
  Bernardis}, {Delabrouille}, {Delouis}, {D{\'e}sert}, {Dolag}, {Dole},
  {Dor{\'e}}, {Douspis}, {Efstathiou}, {Eng}, {Filliard}, {Forni}, {Fosalba},
  {Fourmond}, {Ganga}, {Giard}, {Girard}, {Giraud-H{\'e}raud}, {Gispert},
  {G{\'o}rski}, {Gratton}, {Griffin}, {Guyot}, {Haissinski}, {Harrison},
  {Helou}, {Henrot-Versill{\'e}}, {Hern{\'a}ndez-Monteagudo}, {Hildebrandt},
  {Hills}, {Hivon}, {Hobson}, {Holmes}, {Huffenberger}, {Jaffe}, {Jones},
  {Kaplan}, {Kneissl}, {Knox}, {Lagache}, {Lamarre}, {Lami}, {Lange},
  {Lasenby}, {Lavabre}, {Lawrence}, {Leriche}, {Leroy}, {Longval},
  {Mac{\'{\i}}as-P{\'e}rez}, {Maciaszek}, {MacTavish}, {Maffei}, {Mandolesi},
  {Mann}, {Mansoux}, {Masi}, {Matsumura}, {McGehee}, {Melin}, {Mercier},
  {Miville-Desch{\^e}nes}, {Moneti}, {Montier}, {Mortlock}, {Murphy}, {Nati},
  {Netterfield}, {N{\o}rgaard-Nielsen}, {North}, {Noviello}, {Novikov},
  {Osborne}, {Paine}, {Pajot}, {Patanchon}, {Peacocke}, {Pearson}, {Perdereau},
  {Perotto}, {Piacentini}, {Piat}, {Plaszczynski}, {Pointecouteau}, {Pons},
  {Ponthieu}, {Pr{\'e}zeau}, {Prunet}, {Puget}, {Reach}, {Renault},
  {Ristorcelli}, {Rocha}, {Rosset}, {Roudier}, {Rowan-Robinson}, {Rusholme},
  {Santos}, {Savini}, {Schaefer}, {Shellard}, {Spencer}, {Starck}, {Stassi},
  {Stolyarov}, {Stompor}, {Sudiwala}, {Sunyaev}, {Sygnet}, {Tauber}, {Thum},
  {Torre}, {Touze}, {Tristram}, {van Leeuwen}, {Vibert}, {Vibert}, {Wade},
  {Wandelt}, {White}, {Wiesemeyer}, {Woodcraft}, {Yurchenko}, {Yvon}, \&
  {Zacchei}}]{Planck2011perf}
{Planck HFI Core Team}, {Ade}, P.~A.~R., {Aghanim}, N., {et~al.} 2011, \aap,
  536, A4

\bibitem[{{Roesch} {et~al.}(2012){Roesch}, {Benoit}, {Bideaud}, {Boudou},
  {Calvo}, {Cruciani}, {Doyle}, {Leduc}, {Monfardini}, {Swenson}, {Leclercq},
  {Mauskopf}, {Schuster}, \& {for the NIKA collaboration}}]{roesch}
{Roesch}, M., {Benoit}, A., {Bideaud}, A., {et~al.} 2012, ArXiv e-prints

\bibitem[{{Rubi{\~n}o-Mart{\'{\i}}n} \& {COrE+ Collaboration}(2015)}]{core}
{Rubi{\~n}o-Mart{\'{\i}}n}, J.~A. \& {COrE+ Collaboration}. 2015, in Highlights
  of Spanish Astrophysics VIII, ed. A.~J. {Cenarro}, F.~{Figueras},
  C.~{Hern{\'a}ndez-Monteagudo}, J.~{Trujillo Bueno}, \& L.~{Valdivielso}, 334

\bibitem[{{The COrE Collaboration} {et~al.}(2011){The COrE Collaboration},
  {Armitage-Caplan}, {Avillez}, {Barbosa}, {Banday}, {Bartolo}, {Battye},
  {Bernard}, {de Bernardis}, {Basak}, {Bersanelli}, {Bielewicz}, {Bonaldi},
  {Bucher}, {Bouchet}, {Boulanger}, {Burigana}, {Camus}, {Challinor},
  {Chongchitnan}, {Clements}, {Colafrancesco}, {Delabrouille}, {De Petris}, {De
  Zotti}, {Dickinson}, {Dunkley}, {Ensslin}, {Fergusson}, {Ferreira},
  {Ferriere}, {Finelli}, {Galli}, {Garcia-Bellido}, {Gauthier}, {Haverkorn},
  {Hindmarsh}, {Jaffe}, {Kunz}, {Lesgourgues}, {Liddle}, {Liguori},
  {Lopez-Caniego}, {Maffei}, {Marchegiani}, {Martinez-Gonzalez}, {Masi},
  {Mauskopf}, {Matarrese}, {Melchiorri}, {Mukherjee}, {Nati}, {Natoli},
  {Negrello}, {Pagano}, {Paoletti}, {Peacocke}, {Peiris}, {Perroto},
  {Piacentini}, {Piat}, {Piccirillo}, {Pisano}, {Ponthieu}, {Rath},
  {Ricciardi}, {Rubino Martin}, {Salatino}, {Shellard}, {Stompor},
  {Urrestilla}, {Van Tent}, {Verde}, {Wandelt}, \&
  {Withington}}]{2011arXiv1102.2181T}
{The COrE Collaboration}, {Armitage-Caplan}, C., {Avillez}, M., {et~al.} 2011,
  ArXiv e-prints

\bibitem[{{The Polarbear Collaboration: P.~A.~R.~Ade} {et~al.}(2014){The
  Polarbear Collaboration: P.~A.~R.~Ade}, {Akiba}, {Anthony}, {Arnold},
  {Atlas}, {Barron}, {Boettger}, {Borrill}, {Chapman}, {Chinone}, {Dobbs},
  {Elleflot}, {Errard}, {Fabbian}, {Feng}, {Flanigan}, {Gilbert}, {Grainger},
  {Halverson}, {Hasegawa}, {Hattori}, {Hazumi}, {Holzapfel}, {Hori}, {Howard},
  {Hyland}, {Inoue}, {Jaehnig}, {Jaffe}, {Keating}, {Kermish}, {Keskitalo},
  {Kisner}, {Le Jeune}, {Lee}, {Leitch}, {Linder}, {Lungu}, {Matsuda},
  {Matsumura}, {Meng}, {Miller}, {Morii}, {Moyerman}, {Myers}, {Navaroli},
  {Nishino}, {Orlando}, {Paar}, {Peloton}, {Poletti}, {Quealy}, {Rebeiz},
  {Reichardt}, {Richards}, {Ross}, {Schanning}, {Schenck}, {Sherwin},
  {Shimizu}, {Shimmin}, {Shimon}, {Siritanasak}, {Smecher}, {Spieler},
  {Stebor}, {Steinbach}, {Stompor}, {Suzuki}, {Takakura}, {Tomaru}, {Wilson},
  {Yadav}, \& {Zahn}}]{2014ApJ...794..171T}
{The Polarbear Collaboration: P.~A.~R.~Ade}, {Akiba}, Y., {Anthony}, A.~E.,
  {et~al.} 2014, \apj, 794, 171

\bibitem[{Zmuidzinas(2012)}]{zmudinas}
Zmuidzinas, J. 2012, Annual Review of Condensed Matter Physics, 3, 169

\end{thebibliography}

\end{document}